\documentclass[showpacs,pre,twocolumn,superscriptaddress,longbibliography,amsmath,amssymb]{revtex4-1}

\usepackage{graphicx}
\usepackage{dcolumn}
\usepackage{bm}
\usepackage{amsmath}
\usepackage{color}

\def\N{\mathcal{N}}
\def\V{\mathcal{V}}
\def\l{\mathbf{s}}
\def\q{\mathbf{q}}
\def\er{Erd\H{o}s-R\'enyi }

\def\ie{i.e.}

\def\d{\mathrm{d}}

\begin{document}


\title{Cycles and Clustering  in Multiplex Networks}

\author{Gareth J.~Baxter}
\email{gjbaxter@ua.pt}
\affiliation{Department of Physics \& I3N, University of Aveiro,
  Portugal}
\author{Davide Cellai}
\affiliation{Idiro Analytics, Clarendon House, 39 Clarendon Street, Dublin 2, Ireland}
\affiliation{MACSI, Department of Mathematics and Statistics, University of Limerick, Ireland}
\author{Sergey N. Dorogovtsev}
\affiliation{Department of Physics \& I3N, University of Aveiro,
  Portugal}
\affiliation{A. F. Ioffe Physico-Technical Institute, 194021
  St. Petersburg, Russia}
\author{Jos\'e F. F. Mendes}
\affiliation{Department of Physics \& I3N, University of Aveiro,
  Portugal}

\date{\today}

\begin{abstract}
In multiplex networks, cycles cannot be characterized only by their length, as edges may occur in different layers in different combinations. We define a classification of cycles by the number of edges in each layer and the number of switches between layers. We calculate the expected number of cycles of each type in the configuration model of a large sparse multiplex network. Our method accounts for the full degree distribution including correlations between degrees in different layers. 
In particular, we obtain the numbers of cycles of length $3$ of all possible types. 
Using these, we give a complete set of clustering coefficients and their expected values.
We show that correlations between the degrees of a vertex in different layers strongly affect the number of cycles of a given type, and the number of switches between layers. Both increase with assortative correlations and are strongly decreased by disassortative correlations.
The effect of correlations on clustering coefficients is equally pronounced.
\end{abstract}


\maketitle
\section{Introduction}
The realization that many complex systems cannot be understood by
representing them as a single network, has led to an explosion of
interest in multilayer and multiplex (multiple types of edges)
networks. Applications range from infrastructure \cite{rinaldi2001identifying},
financial \cite{huang2013cascading}, transport \cite{dedomenico2014} and
ecological \cite{pocock2012robustness}.

To properly study multilayer systems, it is essential to understand the fundamental
properties of such structures.
Many 
concepts from single layer networks have already been generalized to multiple layers,
from the degree distribution, to connectivity, adjacency and Laplacian matrices, centrality measures and so on \cite{boccaletti2014structure,kivela2014,dedomenico2016physics}. 
In many cases, the generalization of concepts from single networks---for example, the meaning of ``giant connected component'' \cite{baxter2012avalanche,baxter2015unified}---is not straightforward, and introduces a new dimension to the problem.

In this Paper, we give an analytical description of the statistics of cycles in multiplex networks.
In particular, we consider multiplex networks characterized by a given multi-degree distribution (configuration model).
As this model is the starting point of any network analysis of a real-world system, it is easy to see how important it is to analytically characterize a structural property like the statistics of cycles.
In a multilayer network, the possibility to switch between
layers greatly increases the number of cycle types with respect to the single layer case.
Cycles are then
no longer defined simply by their length. We must take into account
the proportion of the cycle in each layer, as well as the number of
switches between layers. 
In particular, this leads to a set of different 
clustering coefficients generated by different cycles of length $3$. 

Even when two cycles contain the same number of
edges of each color, they can differ in the way colored edges are arranged within the cycle. 
We introduce a matrix $\l$ characterizing the number of
edges of each color, and the number of switches between layers of a
cycle. We give formul{\ae} for calculating the mean number of cycles
corresponding to a given $\l$ in a random graph with a given size and degree distribution.
 As examples, we
calculate the distribution of edge colors and switches in cycles of a
given length, show the effects of degree correlations between layers,
and examine the special case of cycles of length $3$, which allows the
calculation of the generalized clustering coefficients in multiplex
networks.

The statistics of cycles is relevant both from a theoretical and an applicative point of view.
From a theoretical perspective, 
it allows one to understand whether the distribution of cycles
observed in a real world network is significantly different from that
in a random graph with similar statistics \cite{cozzo2015structure}.
Even in the single layer case, the high concentration of finite cycles in real-world networks has been a formidable barrier to analytic treatment, as mathematical models of large networks are typically based on the local tree-like assumption, {\ie} the vanishing of density of finite cycles as the size of the network diverges \cite{newman2003, bianconi2003number,bollobas2003mathematical}.
On the other hand, some 
analytic theories have been quite successful even in real-world networks \cite{melnik2011}, 
suggesting that the role played by the detailed architecture of topological correlations and cycles is not easily characterized.
In multiplex networks, further types of correlations arise naturally, with implications for the structural properties of the multiplex \cite{kim2013,hackett2016}. It is therefore necessary to turn now to multiplex cycles and investigate the statistics of each category based on edge color and layer switches, and the effects of correlations on them.
The properties of cycles is also an important topic of graph theory \cite{bollobas2001book,harary1971}.
A relatively smaller volume of work has been devoted to cycles on colored edge graphs, mainly involving only theorems of existence \cite{abouelaoualim2010}.

The statistics of cycles in multiplex networks is also relevant for a number of applications.
In information technology, the presence of multiple paths of heterogeneous colors is a structural property that improves the robustness of a network \cite{wang2011} and the security of a wireless sensor network from a malicious attack \cite{jain2004}.
Multiplex cycles are also relevant when examining
commuter behavior on multiple transport networks, as they provide alternative routes \cite{dedomenico2014}. 
In such applications, switching between layers may have a time or monetary cost, 
a consideration which is absent in single layer networks. The statistics of switches is  therefore an essential part of any analysis of cycles in multiplex networks.

This paper is organized as follows.
In Section II, we describe a classification of cycles in multiplex networks and explain our formula to calculate the number of cycles within a given class.
In Section III, we use our formula to calculate multiplex clustering coefficients.
Finally, in Section IV we state our conclusions.

\section{Statistics of cycles}

\begin{figure}[htb]
	\includegraphics[width=\columnwidth]{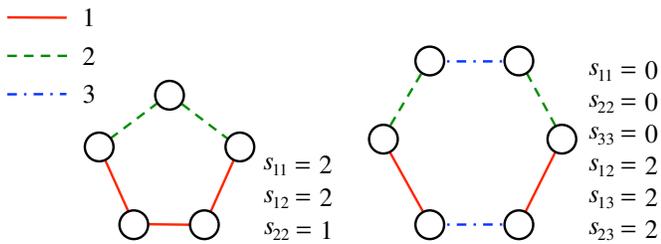}
	\caption{Notation of multiplex cycles. }
	\label{fig:cycle-notation}
\end{figure}
We characterize a given
cycle in a multiplex network, by the matrix 
$\l = \{s_{ab}\}_{a,b=1,\dots,M}$, where
each element $s_{ab}$ defines the number of nodes in the cycle which 
connect an edge of type $a$ with an edge of type $b$. 
When $a=b$, this counts the number of nodes where the cycle remains 
in the same layer. 
When $a\neq b$, $s_{ab}$ counts the number of switches between layer $a$ and layer $b$. 
In this Paper, we identify and study the classes of equivalence of cycles with the same $\l$.
In addition, here we only consider cycles without an orientation, therefore 
the order of the switches  is not important, {\ie} $s_{ba} = s_{ab}$.
Examples of cycles with a given $\l$ are shown in Fig.~\ref{fig:cycle-notation}.
As a consequence of this definition,
the total number of edges in layer $a$ is then $n_a = s_{aa} + \frac{1}{2}\sum_{b\neq a} s_{ab}$. 
Clearly, the number of switches must be such that $n_a$ is an integer for all $a$. The total length of a cycle $L$ is the sum of all entries of $\l$. 

We consider a generalization of the configuration model to multiplex networks, {\ie} large sparse random multiplex networks with $N$ nodes in $M$ layers, defined by
the joint multidegree distribution $P(q_1,q_2, ...,q_M)$. 
This ensemble includes all possible configurations with multidegree sequence sampled from this distribution with equal statistical weight \cite{bender1978asymptotic,bollobas1980probabilistic}.

To calculate the mean number of cycles $\N(\l)$ with a given $\l$ in a random graph,
we first count the number of ways we can select, from the given multidegree distribution,
the nodes that have the connectivity required for each $s_{ab}$; 
then, we count the number of ways we can
connect these nodes to form a cycle. 
This method is similar to that
used in, for example, Ref. \cite{bianconi2005loops}
but we extend it to account for edges of different colors and switches
between layers.
The results can be written as the product of several factors:
\begin{equation}\label{N_general}
\N(\l) = G(\l)  W(N,\l) R[N,\l, P(\q)]\,,
\end{equation}
where
$R(N,\l,P(\q))$ counts the number of ways one can select pairs of edges connected to $L$ nodes, 
in the correct numbers to match
the elements
of matrix $\l$,
$W(N,\l)$ counts the number of graphs in the ensemble containing the
cycle, and
$G(\l)$ counts the number of ways of arranging the selected nodes to
form a cycle.
In principle one can complete this calculation for an arbitrary cycle
in a multiplex with an arbitrary number of layers. However, the
calculation of $G(\l)$ becomes somewhat complicated for more than two
layers, for anything but the shortest cycles.

Let us focus, now, on a two layer random multiplex (duplex), defined by the
joint degree distribution $P(q_1,q_2)$.
In the case of two layers, $\l$ has three entries: $s_{11}$, $s_{22}$,
and $s_{12}$. For a given $L$, $s_{11}$ and $s_{22}$ can take any
value from $0$ to $L$, while the number of switches $s_{12} \equiv
2p$, for integer $p$, to ensure that the number of edges of each color
is integral, while all three must satisfy $s_{11} + s_{22} + s_{12} =
L$.
The formula (\ref{N_general}) can be calculated explicitly in the
asymptotic case characterized by $L\ll N$.

Without switches, $G(\l)$ is simply the number of possible orderings of the 
$L$ nodes, dividing by $2L$ as each direction and starting point in the ordering is equivalent:
\begin{equation}\label{G_1layer}
G(L,0,0) = \frac{L!}{2L}
\end{equation}
and similarly for $G(0,L,0)$. 
When $s_{12} > 0$, $G(\l)$ is given by the number of ways of ordering the 
switches $(s_{12}-1)!$ multiplied by the number of ways $D(s_{11},p)D(s_{22},p)$ 
of placing the non-switching nodes in the spaces between the switches, where
\begin{eqnarray}
	D(s,p) &=&  \sum_{n_1=0}^{l} {s\choose n_1}n_1! \sum_{n_2=0}^{s-n_1} 
	{s-n_1\choose n_2}n_2\dots \nonumber\\
	&&\dots \sum_{n_{p-1}=0}^{s-\sum_{j=1}^{p-2}n_j} 
	{s-\sum_{j=1}^{p-2}n_j\choose n_{p-1}}n_{p-1}! \left[s-\sum_{j=1}^{p-1}n_j\right] \nonumber\\
	&=& \frac{(s+p-1)!}{(p-1)!}
	\end{eqnarray}
Thus
\begin{equation}\label{G}
	G(s_{11},s_{22},s_{12}) = \left\{%
		\begin{array}{lcrcl}
			(s_{12}-1)! D(s_{11},p) D(s_{22},p) && s_{12}>0\\
			\frac{(s_{11}+s_{22})!}{2(s_{11}+s_{22})} && s_{12}=0\\
		\end{array}
	\right..
\end{equation}

The number of ways to form layer $1$ is the number of ways to connect $c_1N$ stubs in pairs: $(c_1N-1)(c_1N-3)...3\cdot1 = (c_1N-1)!!$, while 
the number of ways to connect the edges not forming part of the loop is  $(c_1N-2s_{11}-s_{12}-1)!!$.
Hence the fraction of layer $1$ configurations containing the loop is the ratio of these two numbers. Repeating for layer $2$, we find that 
$W(N,\l)$ can be simply written
\begin{equation}\label{W_main}
	W(N,\l) = \frac{(c_1N\!-\!2s_{11}\!-\!s_{12}\!-\!1)!! (c_2N\!-\!2s_{22}\!-\!s_{12}\!-\!1)!!}
	{(c_1N-1)!! (c_2N-1)!!}.
\end{equation}

\subsection{Formul{\ae} for short cycles}

When $L \ll N$, 
the factor $W(N,\l)$ 
can be approximated as
\begin{equation}\label{W_approx}
W(N,\l) = \frac{1}{N^L \langle q_1\rangle^{s_{11}+p} \langle q_2\rangle^{s_{22}+p}}.
\end{equation}
Furthermore,
we can treat the selection of nodes as being done with replacement, meaning 
that $R(N,\l,P(\q))$ can be simply written as a product of terms for each node in the cycle:
\begin{align}
	R(N,\l,P(\q)) 
	= 
	\frac{ N^L\langle q_1(q_1-1)\rangle^{s_{11}}
\langle q_2(q_2-1)\rangle^{s_{22}}
\langle q_1q_2\rangle^{s_{12}}}
{s_{11}!s_{22}!s_{12}!} \label{R_short}
\end{align}
where $\langle ...\rangle$ indicates averages with respect to the
degree distribution (ensemble averages).

When Eq. (\ref{R_short}) is combined with $G(\l)$ and $W(N,\l)$ as given by Eqs. (\ref{G}) and (\ref{W_approx}) 
we 
find that, when none of $s_{11}, s_{22}, s_{12}$ is zero,
\begin{align}\label{N_full}
  \N(\l) &=
\binom{s_{11}+p-1}{s_{11}}\binom{s_{22}+p-1}{s_{22}}\frac{1}{2p}
  \left[\frac{\langle  q_1(q_1-1)\rangle}{\langle q_1\rangle}\right]^{s_{11}}\nonumber\\
  \times
&  \left[\frac{\langle  q_2(q_2-1)\rangle}{\langle  q_2\rangle}\right]^{s_{22}}
  \left[\frac{\langle  q_1q_2\rangle}{\sqrt{\langle q_1\rangle \langle  q_2\rangle}}\right]^{2p}\,,
\end{align}

In the case $s_{12} = 0$, the cycle consists of only one color, so $\N(\l) = 0$ unless either $s_{22} = 0$ or $s_{11} = 0$, in which case
\begin{equation}
	\N(L,0,0) 
	 =  \frac{1}{2L} \left[\frac{\langle
             q_1(q_1-1)\rangle}{\langle q_1\rangle}\right]^{L} .
       \label{N_single_color}
\end{equation}
This coincides with the single layer result found in, for example, Ref. \cite{bianconi2005loops}.
Similarly, the formula for $\N(0,L,0)$ is found simply by exchanging the subscripts. 

On the other hand, when $s_{11} = 0$, it is still possible to have $s_{12}>0$, when each segment in layer $1$ consists of only a single edge (i.e. each switch between layers is immediately followed by another switch). Then 
\begin{multline}\label{N_no_l1}
 	\N(0,s_{22},2p) =\\
	\binom{s_{22}+p-1}{s_{22}}\frac{1}{2p}
  \left[\frac{\langle  q_2(q_2-1)\rangle}{\langle  q_2\rangle}\right]^{s_{22}}
  \left[\frac{\langle  q_1q_2\rangle}{\sqrt{\langle q_1\rangle \langle  q_2\rangle}}\right]^{2p}\,,
\end{multline}
and similarly for the case $s_{22} = 0$ but $s_{12}>0$ and $s_{11}>0$, by exchanging the subscripts $1$ and $2$.
If we project the two layers onto a single network, we recover the existing result for a single colored network, which has the same form as Eq. (\ref{N_single_color}).

These results are valid for $L \ll N$, such as in the limit $N\to \infty$. The expected number of cycles in other cases, for longer cycles (when terms of $\mathcal{O}(1/N)$ can't be neglected) can be found by a more precise derivation, which we outline in the Appendix. 

\subsection{Representative examples}
\begin{figure}[htb]
\includegraphics[width=\columnwidth]{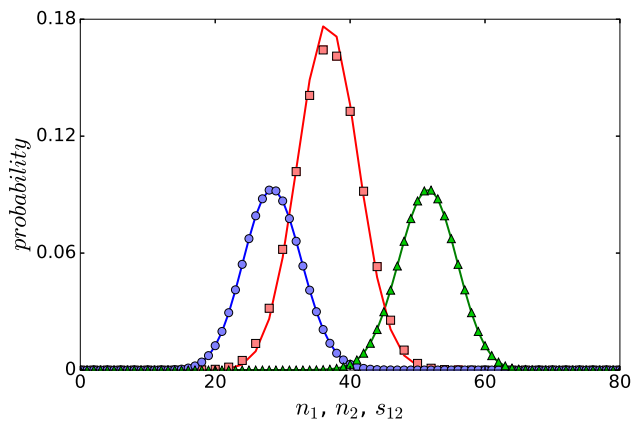}
\caption{
Distribution of number of edges of type $1$ (circles, blue online) and $2$ (triangles, green online) and number of switches $s_{12}$ (squares, red online) in cycles of length $L = 80$ in two uncorrelated layers with Poisson degree distributions having mean degrees $\langle q_1\rangle = 25$ and $\langle q_2\rangle = 45$. 
Symbols are from summation of Eq. (\ref{N_full}), solid lines are binomial distributions for $L$ trials with probabilities $\langle q_1\rangle/(\langle q_1\rangle+\langle q_2\rangle)$, $\langle q_2\rangle/(\langle q_1\rangle+\langle q_2\rangle)$, and $\langle q_1\rangle\langle q_2\rangle/(\langle q_1\rangle+\langle q_2\rangle)^2$ respectively.}
\label{fig:distributions}
\end{figure}
The number of cycles having exactly $n_1 = s_{11} + \frac{1}{2}s_{12}$ edges of type $1$ for a fixed $L$ can be found by summing Eq. (\ref{N_full}) over each $s_{ab}$.
In the absence of inter-layer degree correlations, the resulting distributions for $n_1$ match the Binomial distribution found by selecting $L$ edges at random from the network, as  shown in Fig. \ref{fig:distributions}. The mean number of type $1$ edges is $p_1 = \langle q_1\rangle / (\langle q_1\rangle + \langle q_2 \rangle)$, and similarly for $n_2$. 
In addition, the number of switches $s_{12}$, which must always be even, can be found by summing over $s_{11}$ and $s_{22}$. The mean number of switches $\langle s_{12}\rangle$ is well predicted by $2L\langle q_1\rangle \langle q_2\rangle / (\langle q_1\rangle + \langle q_2 \rangle)^2$, which is the expected number of mismatches when pairing $L$ randomly chosen edges, and the distribution is also well matched by a binomial distribution.

The number of cycles for a given matrix $\l$ in Eq. (\ref{N_full}) depends only on the moments of the joint degree distribution. This means that networks 
may have quite different degree distributions, and hence 
different structures, but if the relevant moments are the same, so will be the average number of cycles of each kind.

\begin{figure}[htb]
\includegraphics[width=0.48\columnwidth]{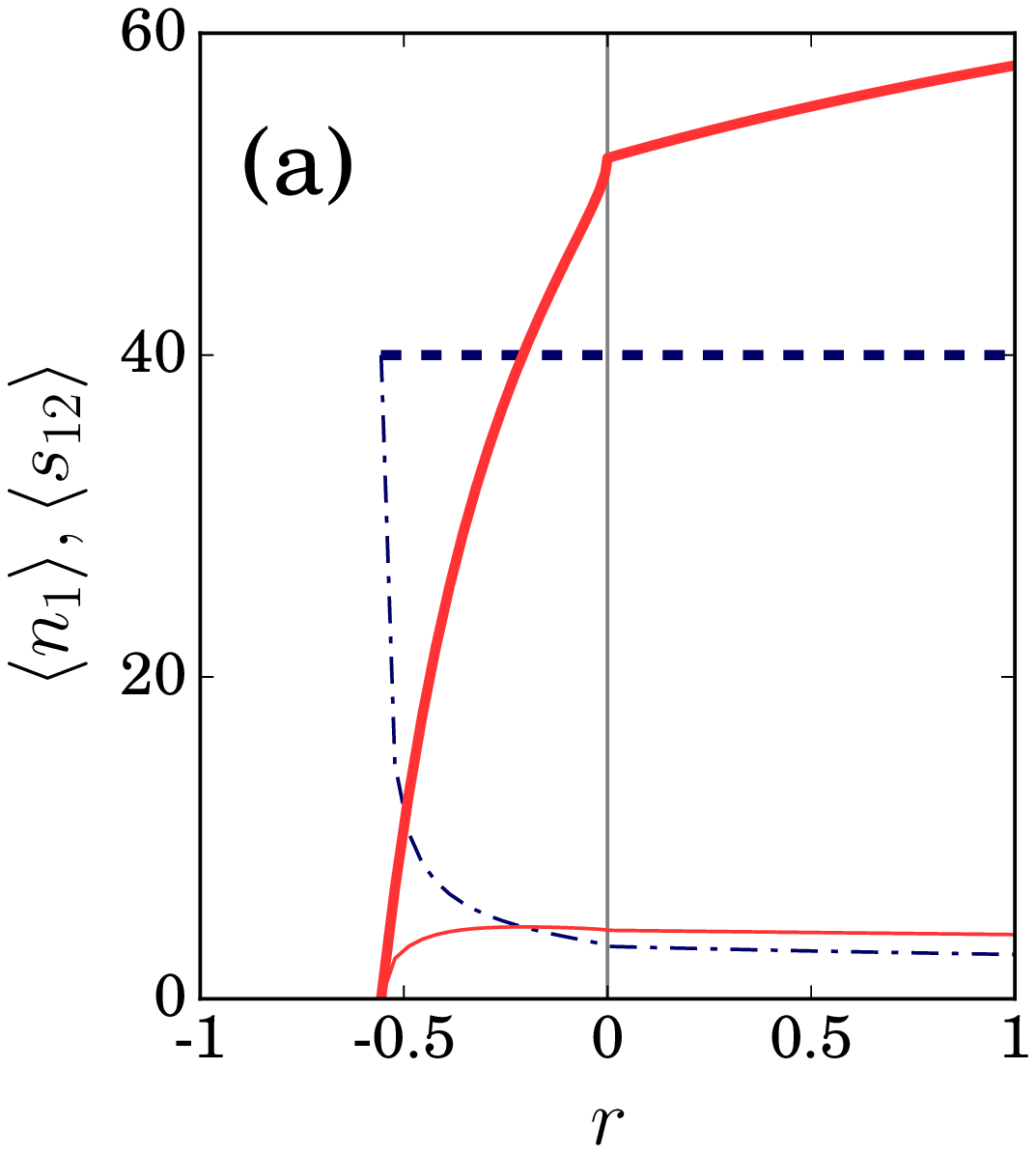}
\includegraphics[width=0.48\columnwidth]{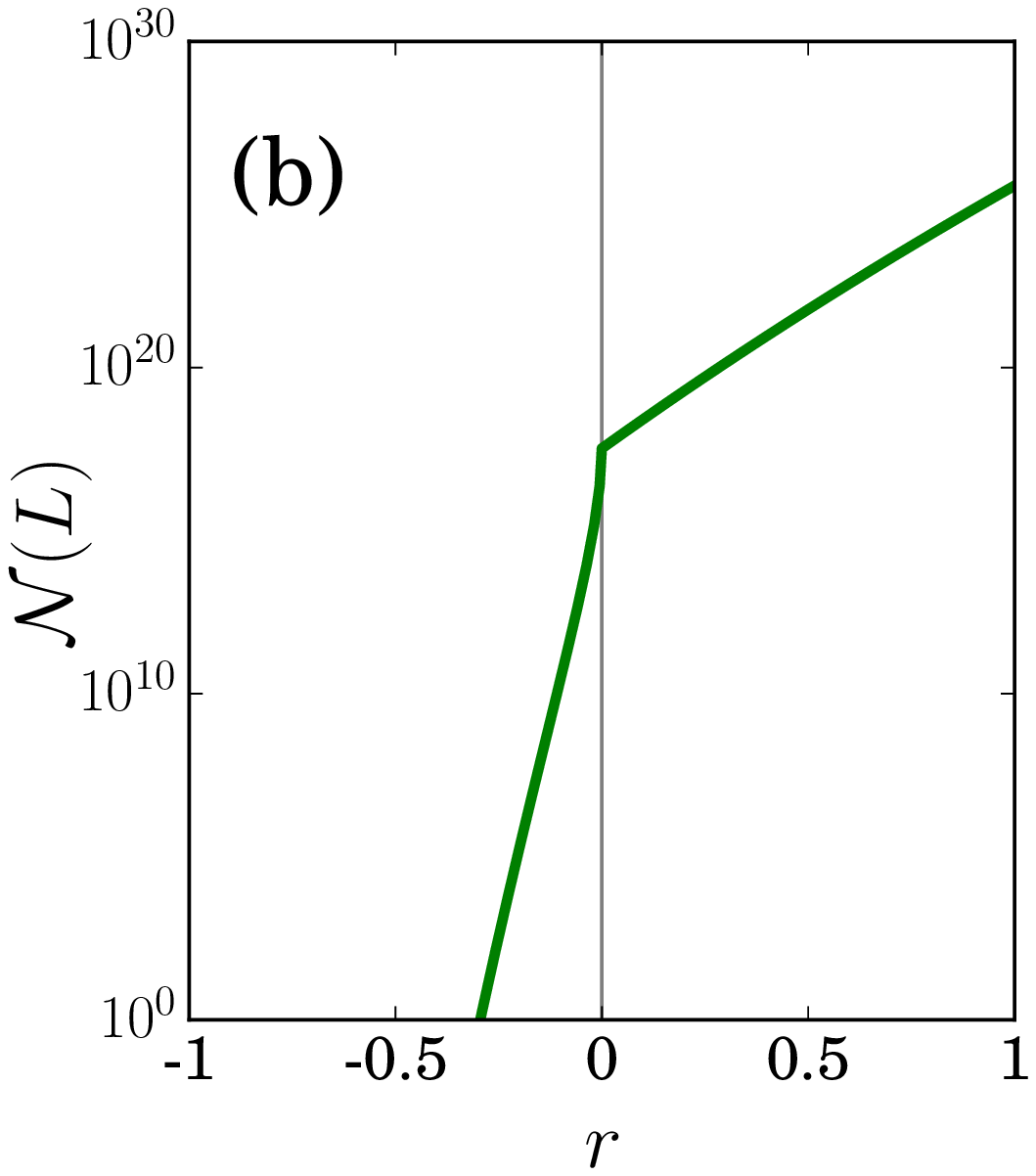}
\caption{(a) Mean for $n_1$ (blue, dashed) and $s_{12}$ (red, heavy solid) and corresponding standard deviations (blue dash-dot, red light, respectively) as a function of Pearson correlation coefficient $r$ for the degrees of a vertex in different layers. Assortative correlations ($r<0$) are created using a joint degree distribution of the form
$P(q_1,q_2)  = \rho\frac{1}{2}[(P(q_1)+P(q_2)]\delta_{q_1,q_2} + (1-\rho)P(q_1)P(q_2)$.  disassortative correlations ($r<0$) are of the form $P(q_1,q_2)  = \rho[(P(q_1)\delta_{q_2,0}+P(q_2)\delta_{q_1,0}] + (1-\rho)P(q_1)P(q_2)$, where $P(q) \propto q^{-\gamma}$, with $\gamma=3.7$, is a power-law distribution, and $L = 80$. Maximum disassortative correlation occurs when the two layers no longer overlap, at which point $n_1=0$. The corresponding value of $r$ depends on $\gamma$. Figures are qualitatively similar for any value of $\gamma > 3$. 
(b) Total number of cycles as a function of $r$.
}\label{fig:correlation_number}\label{fig:correlation_means}
\end{figure}

In simplex networks, it has been shown that degree-degree correlations for neighboring vertices affect quite deeply the number of cycles \cite{bianconi2006}.
In multiplex networks, it is natural to ask about the effect of degree correlations across layers.
Indeed, we can see from the last term in Eq. (\ref{N_full}) that interlayer degree correlations affect the number of cycles having a given number of switches between layers. For a given $L$, assortative interlayer degree correlations will tend to increase the number of switches, as high degree nodes in one layer, which are more frequently visited, will also have more available edges in the other layers. 
Conversely, disassortative correlations will tend to decrease the number of switches. The mean number of edges of type $1$ remains constant, although the distribution changes. These effects can be seen in Fig. \ref{fig:correlation_means}, in which we plot the mean and standard deviation of $n_1$ and $s_{12}$ as a function of the Pearson correlation coefficient $r$ for the degrees of a vertex in different layers \cite{newman2003}.
We see that in the extreme case of disassortative correlations, there are no switches and the entire cycle is of one color, as the standard deviation of $n_1$ reaches the maximum value of $L/2$. An even more dramatic change is seen in the total number of cycles of a given length. The right panel in Fig.~\ref{fig:correlation_number} shows the total number of cycles of length $L$ as a function of $r$. Disassortative correlations greatly restrict the possible number cycles that can be formed, while assortative correlations greatly increase it.
Note that the apparently sharp inflections at $r=0$ result simply from the use of different functional forms for assortative and disassortative correlations.

\section{Clustering}

The clustering coefficient of a network is related to the number of
triangles, that is, cycles of length three.
In such short cycles, the computation of factor $G(\l)$ is
straightforward, thus we can calculate the number of
cycles of length $3$ for any number of layers. Such a cycle may be
entirely within one layer: $\N_{m}^{(1)} = z_m^3/6c_m^3$; have two
edges in one layer ($m$) and one edge in a second layer ($n$): $\N_{m,n}^{(2)} =
z_m \langle q_mq_n\rangle^2/ 2 c_m^2c_n$; or have one edge each in
three different layers: $\N_{m,n,r}^{(3)} = \langle q_mq_n\rangle
\langle q_mq_r\rangle \langle q_nq_r\rangle / c_mc_nc_r$
, where $z_m \equiv \langle q_m(q_m-1)\rangle$ and $c_m = \langle
q_m\rangle$.
We can then define a global clustering coefficient by
\begin{align}
  C &= \frac{3\sum_{\l : L=3} \N(\l)}{\V(N)}
      \nonumber \\  
  &= \frac{ 3{\displaystyle \sum_m} \N_{m}^{(1)}
          + 3{\displaystyle \sum_{m, n\neq m}} \N_{m,n}^{(2)}
      + 3{\displaystyle \sum_{\substack{m,n\neq m\\ r \neq m,n}} }\N_{n,m,r}^{(3)} }
  {N\left[ \sum_m z_m + 2\sum_{m,n\neq m} \langle
      q_mq_n\rangle\right]}\,,
  \label{C_M}
\end{align}
where $\V(N)$ is the number of adjacent edge-pairs in the graph, and the
summation is over all cycle matrices $\l$ having length $L=3$.

One may also define partial clustering coefficients for triangles
entirely within a given  layer, two given layers, or three given
layers ($C_{m,n,r}^{(3)}$) respectively:
\begin{align}
  C_{m}^{(1)} &= \frac{3\N_{m}^{(1)}}{\frac{1}{2}N z_m }
  = \frac{z_m^2}{Nc_m^3 },\\
  C_{m,n}^{(2)} &= \frac{3\N_{m,n}^{(2)}}{\frac{1}{2}N \left[z_m + 2\langle
      q_mq_n\rangle \right] }
  = \frac{3z_m \langle q_mq_n\rangle^{2}}
  {N \left[z_m + 2\langle q_mq_n\rangle \right]c_m^2 c_n},\\
  C_{m,n,r}^{(3)} &= \frac{3\N_{m,n,r}^{(3)} }{\frac{1}{2}N \left[\langle
      q_mq_n\rangle  + \langle  q_mq_r\rangle + \langle  q_nq_r\rangle
      \right] }\\
  &= \frac{6\langle q_mq_n\rangle
\langle q_mq_r\rangle \langle q_nq_r\rangle }
  {N c_mc_nc_r\left[\langle
      q_mq_n\rangle  + \langle  q_mq_r\rangle + \langle  q_nq_r\rangle \right] },
\end{align}
or triangles entirely in one, two, or three layers, regardless of
which particular layers they are (these less fine-grained coefficients were defined and calculated numerically in Ref. \cite{cozzo2015structure}):
\begin{align}
  C^{(1)} 
  &=  \frac{\sum_m (z_m/c_m)^3}{N\sum_m z_m}\,, \label{C1}\\
  C^{(2)} 
  &= \frac{3\sum_{m,n\neq m}z_m \langle q_mq_n\rangle^2/( c_m^2c_n ) }
  { N\sum_{m,n\neq m} \left[z_m + 2 \langle
      q_mq_n\rangle\right] }
  \label{C2}\,, \\
  C^{(3)} 
  &= \frac{6\sum_{m,n\neq m,r \neq m,n}\langle q_mq_n\rangle
\langle q_mq_r\rangle \langle q_nq_r\rangle / c_mc_nc_r }
  { N\sum_{m,n\neq m, r \neq m,n} \langle q_mq_n\rangle }\,. \label{C3}       
\end{align}

These formul{\ae} give the expected clustering coefficients for large
random graphs, taking into account the full degree distribution. This
gives a more accurate result than found by simply matching the mean
degree to an \er network \cite{cozzo2015structure}, for which the clustering
coefficients can be calculated by considering the probability for a
given edge to be present or absent:
\begin{align}
C_{m (ER)}^{(1)} &= \frac{c_m}{N}\,, \label{C_er1}\\
C_{m,n (ER)}^{(2)} &= \frac{3c_mc_n}{N(c_m+2c_n)}\,, \\
C_{m,n,r (ER)}^{(3)} &= \frac{6 c_mc_nc_r }{N (c_mc_n  + c_mc_r + c_nc_r) }\,, \\
C^{(1)}_{ER} &= \frac{\sum_m c_m^3}{N\sum_m c_m^2}\,, \\
C^{(2)}_{ER} &= \frac{3\sum_{m,n\neq m}c_m^2c_n }{N\sum_{m,n\neq m}(c_m^2 + 2c_mc_n)}\,,\\
C^{(3)}_{ER} &= \frac{6\sum_{m,n \neq m, r \neq m,n}c_mc_nc_r}{N\sum_{m,n\neq m, r \neq m,n} c_mc_n}\,, \\
 C_{ER} &= \frac{\sum_m c_m}{2N} ,
 \label{C_er7}
\end{align}
which coincide with the results found by inserting uncorrelated Poisson degree distributions into Eqs. (\ref{C_M})-(\ref{C3}).

\begin{figure}[htb]
\includegraphics[width=0.48\columnwidth]{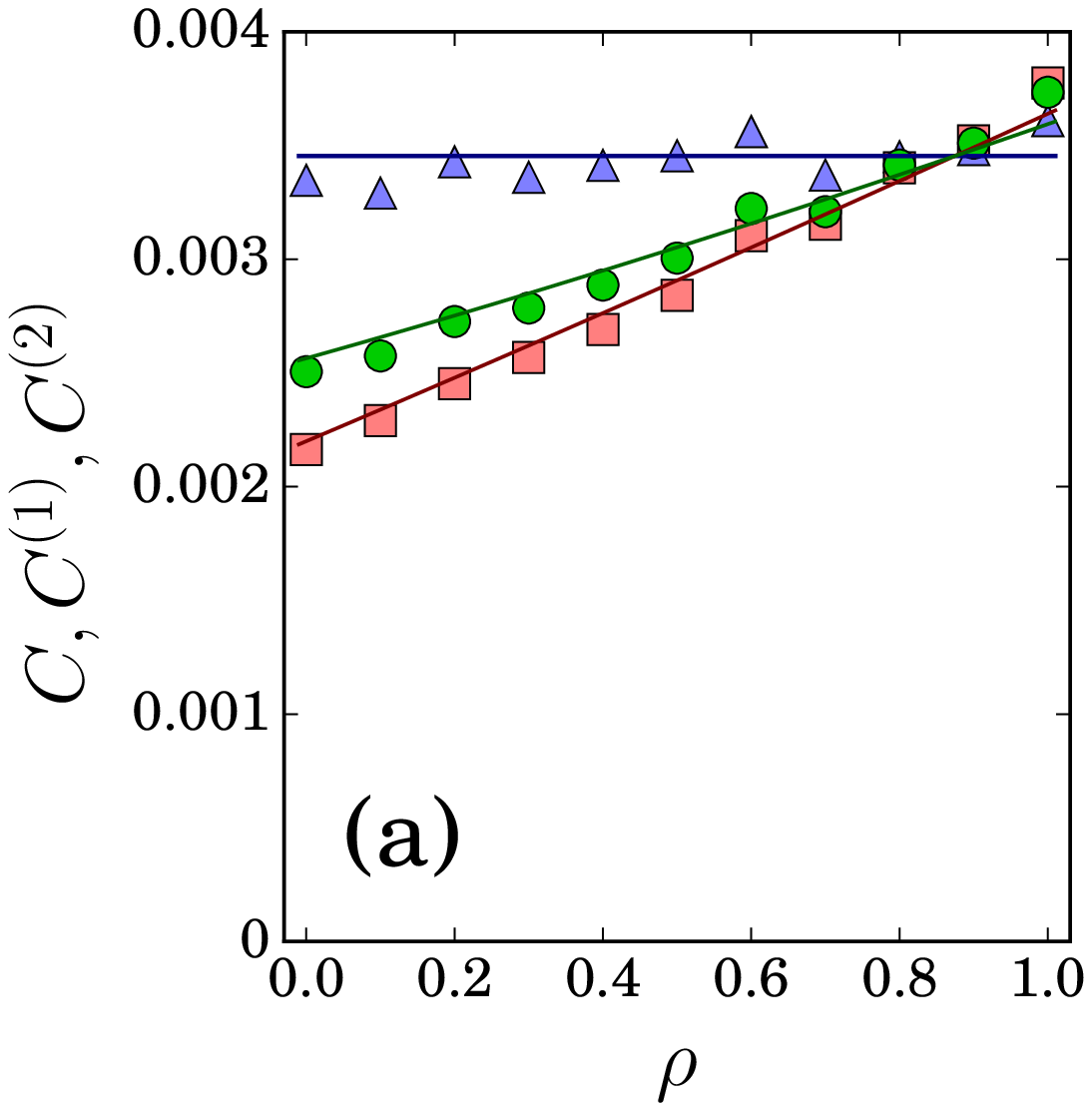}
\includegraphics[width=0.48\columnwidth]{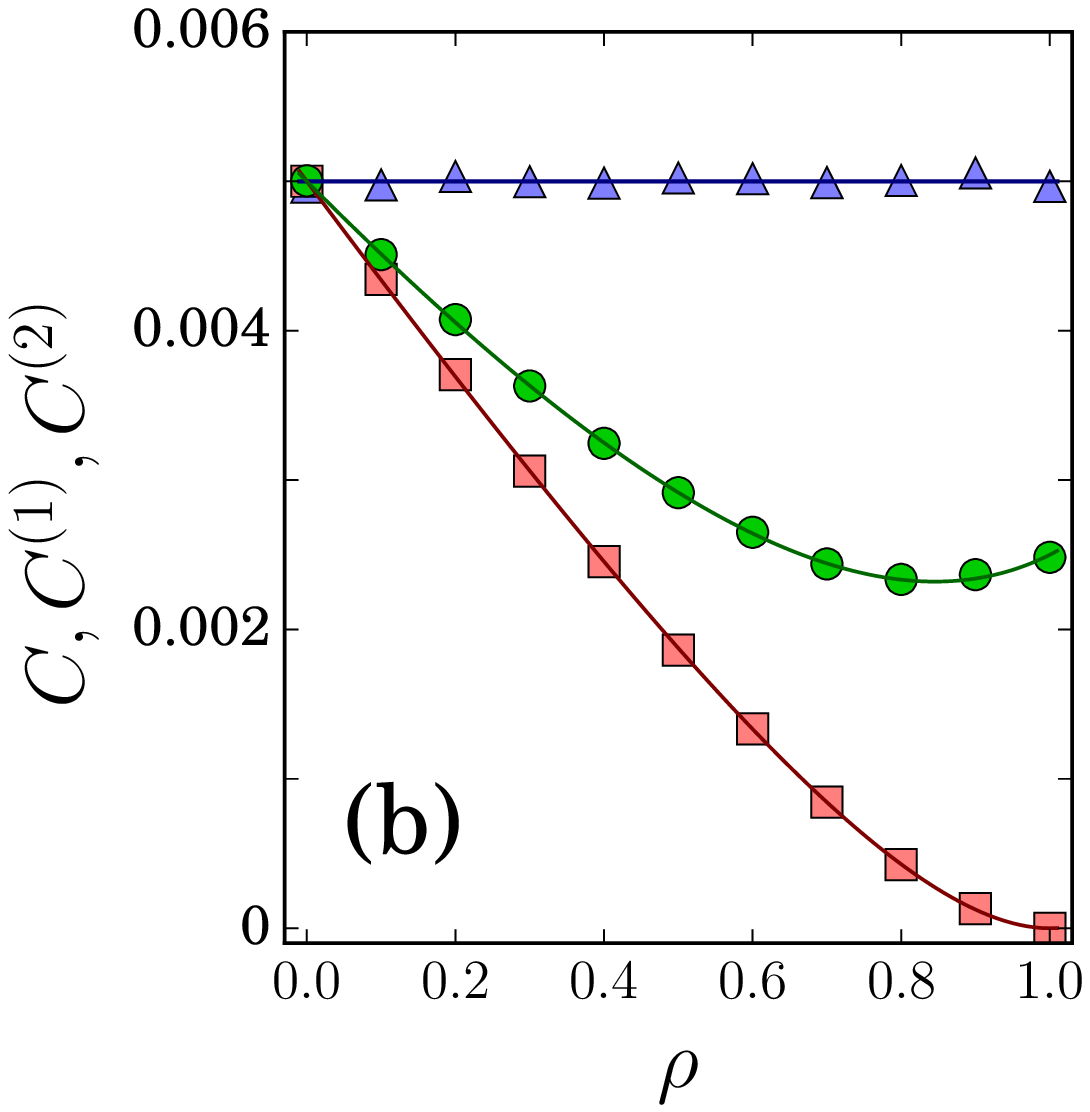}
\caption{(a) Total clustering coefficient, $C$ (circles, green), single layer, $C^{(1)}$ (triangles, blue) and two layer, $C^{(2)}$ (squares, red), clustering coefficients for a two layer multiplex network with assortative inter layer degree correlations, joint degree distribution 
$P(q_1,q_2)  = \rho\frac{1}{2}[(P(q_1)+P(q_2)]\delta_{q_1,q_2} + (1-\rho)P(q_1)P(q_2)$, 
where $P(q) \propto q^{-\gamma}$, with $\gamma=2.9$ for $q \in [10,100]$. Each layer has $N = 10^4$ vertices. For $\rho = 0$ correlations are absent, while for $\rho=1$ degrees are perfectly correlated. Solid lines are expected theoretical values using Eqs. (\ref{C_M}), (\ref{C1}), and (\ref{C2}).
(b) Total $C$, single layer, $C^{(1)}$, and two layer, $C^{(2)}$, clustering coefficients for a two layer multiplex network with disassortative inter layer degree correlations, joint degree distribution 
$P(q_1,q_2)  = \rho[(P(q_1)\delta_{q_2,0}+P(q_2)\delta_{q_1,0}] + (1-\rho)P(q_1)P(q_2)$,
where $P(q)$ is a Poisson distribution with mean $50$. For $\rho = 0$ correlations are absent, while for $\rho=1$ degrees are perfectly disjoint. Solid lines are expected theoretical values using Eqs. (\ref{C_M}), (\ref{C1}), and (\ref{C2}).
}\label{fig:clustering}
\end{figure}

To illustrate the importance of taking correlations into account, we compared our formul{\ae} Eqs.~(\ref{C_M})-(\ref{C3}) with measurements of synthetic networks. The results are summarized in Fig.~\ref{fig:clustering}. This shows that our method successfully accounts for the effects of broad degree distributions, and inter-layer correlations. In comparison, the \er formul{\ae} are generally only accurate for uncorrelated \er layers and fail completely in the presence of strong correlations between layers.

\section{Conclusions}

In single layer networks, cycles are characterized by their length. In multiplex networks, there are many more possibilities.
In particular, there is the possibility to switch between layers, and this must be accounted for.
In this Paper we have introduced a classification for cycles in multiplex networks based on the number of edges in each layer, and the number of switches between layers. We further calculated the expected number of each type of cycle in a large random multiplex.
Our results are valid for any multi-degree distribution, including distributions characterized by arbitrary degree correlations between layers. 
 Interestingly, our formul{\ae} show that the first and second order moments of the multi-degree distribution are sufficient to determine the statistics of cycles in large multiplex networks.
The effect of correlations between a vertex's degrees in different layers affects these statistics through the degree-degree moment $\langle q_mq_n\rangle$. Assortative correlations tend to increase the number of switches in cycles of a given length, while also increasing the total number of cycles. Disassortative correlations, on the other hand, may greatly reduce both the total number of cycles and the total number of switches within these cycles.

These results further allow us to give the expected clustering coefficients in multiplex networks. The possibility that a closed triangle may have edges in multiple layers requires a more detailed discrimination of clustering coefficients.
We give a complete classification of the various possible clustering coefficients and give their expected values. 
Inter layer correlations again have a strong effect, significantly increasing or decreasing the mixed-layer clustering coefficients.
These results give a much more precise view of cycles and clustering in multiplex networks than using the mean degree alone, and establish the proper baseline for comparison with real-world networks.

\begin{acknowledgments}
{We acknowledge useful discussions with Mikko Kivel\"a. 
This work was partially supported by the FET
proactive IP project MULTIPLEX 317532. GJB was supported by
the FCT grant No. SFRH/BPD/74040/2010.
DC was supported by Science Foundation
Ireland, Grant 14/IF/2461.}
\end{acknowledgments}

\appendix*

\section{The factor $R$ for longer cycles}

The derivation given in the main body of this Paper is valid for $L \ll N$. If one wants to consider longer cycles, a more complex derivation is required, which we sketch here.

\subsection{Cycles in only one layer}

For orientation, we first outline the derivation for cycles within a single layer, which is identical to that for a single network, and follows the method used in \cite{bianconi2005loops}.

Consider selecting an arbitrary set of nodes from the network, of which $n_q$ have degree $q$. There are $\binom{N_q}{n_q}$ ways to select $n_q$ nodes, where $N_q = NP(q)$. We sum over all possible such sets and use a delta function to count only sets containing exactly $L$ nodes. There are $q(q-1)$ ways to select the edges which connect a node of degree $q$ in the cycle. Together these considerations mean we can write
\begin{multline}
R(N,L,P(q)) = \\
N^L \!\sum_{\{n_q\}}\! \delta\! \left(\!\sum_q \!n_q\! -\! L\! \right)\! \prod_q \binom{N_q}{n_q}\! \left[ \frac{q(q-1)}{N}\right]^{n_q}.
\end{multline}

We can write the delta function in integral form, and after replacing the sum over sets $\{n_q\}$ with a sum over $n_q$ after the product, we have
 \begin{align}
R&(N,L,P(q)) \nonumber\\
&= \frac{N^L}{2\pi} \int \d x \, 
\prod_q 
\sum_{n_q=0}^{N_q} 
\binom{N_q}{n_q} \left[ \frac{e^{ix}q(q-1)}{N}\right]^{n_q} e^{-ixL}\nonumber\\
 &= 
\frac{N^L}{2\pi} \int \d x \, 
\prod_q \left[ 1+ \frac{e^{ix}q(q-1)}{N} \right]^{N_q} e^{-ixL}.
\end{align}
Writing the product as a sum within an exponential, and 
recognising that as $N_q = NP(q)$, this gives an average over degree, we have
\begin{multline}
R(N,L,P(q)) = \\
\frac{N^L}{2\pi} \int \d x \,
\exp\left\{ -ixL + N \left\langle \log\left[1+ \frac{e^{ix}q(q-1)}{N}\right]\right\rangle \right\}.\label{R1}
\end{multline}

To evaluate this integral, we assume the integrand to be strongly peaked around a certain value $x^*$.
Let
\begin{equation}
\phi(x) = \frac{-ixL}{N} + \left\langle \log\left[1+ \frac{e^{ix}q(q-1)}{N}\right]\right\rangle.\label{phi}
\end{equation}
A local maximum of this function occurs at $x^*$
which is the solution of
\begin{equation}
\left\langle \frac{q(q-1)}{Ne^{-ix^*}+ q(q-1)} \right\rangle = \frac{L}{N}.\label{x_star}
\end{equation}
The second derivative evaluated at $x^*$ is
\begin{equation}
\frac{\d^2 \phi}{\d x^2}\Big|_{x=x^*} = - \left\langle \frac{q(q-1)}{[Ne^{-ix^*}+ q(q-1)]^2} \right\rangle Ne^{-ix^*}.
\end{equation}
These results can be used to replace $\phi(x)$ with a Taylor expansion about $x^*$ in Eq. (\ref{R1}).
It has a simple Gaussian form that can easily be evaluated to give
\begin{multline}
R(N,L,P(q)) =\\ \frac{N^L}{\sqrt{2\pi}} e^{N\phi(x^*)}
\left\{ 
 \left\langle \frac{q(q-1)}{[Ne^{-ix^*}+ q(q-1)]^2} \right\rangle N^2e^{-ix^*}
\right\}^{-1/2}.\label{R1_final}
\end{multline}
To find an explicit expression, we must solve Eq. (\ref{x_star}) and evaluate $\phi(x^*)$ and hence Eq. (\ref{R1_final}).

In the case $L \ll N$, $Ne^{-ix}$ dominates $q(q-1)$, so that Eq. (\ref{x_star}) becomes
\begin{equation}
\langle q(q-1) \rangle = Le^{-ix^*}
\end{equation}
so that 
\begin{equation}
N\phi(x^*) = L + \log\left(\frac{\langle q(q-1)\rangle}{L}\right)^L
\end{equation}
and
\begin{multline}
\left\{ 
 \left\langle \frac{q(q-1)}{[Ne^{-ix^*}+ q(q-1)]^2} \right\rangle N^2e^{-ix^*}
\right\}^{-1/2}
= \frac{1}{\sqrt{L}}
\end{multline}
giving
\begin{equation}
R(N,L,P(q)) = 
\frac{N^L}{\sqrt{2\pi L}} 
e^L \left[\frac{\langle q(q-1)\rangle}{L} \right]^L
=\frac{N^L}{L!} \langle q(q-1)\rangle^L
\end{equation}
which when combined with Eqs. (\ref{G_1layer}) and (\ref{W_approx}) gives exactly Eq. (\ref{N_single_color}).

If instead we expand the lefthand side of Eq. (\ref{x_star}) we find
\begin{equation}
\frac{s}{N} =  \frac{\langle q(q-1)\rangle}{Ne^{-ix^*}}  - \frac{\langle q^2(q-1)^2\rangle}{N^2e^{-2ix^*}} + ... 
\end{equation}
and keeping the two leading terms 
, the solution is
\begin{align}
e^{-ix^*}
&\approx \frac{\langle q(q-1)\rangle}{s} - \frac{\langle q^2(q-1)^2\rangle}{N\langle q(q-1)\rangle}
\end{align}
which can be substituted into Eqs. (\ref{phi}) and (\ref{R1_final}) to find the expected number of cycles when $L$ is not small compared with $N$ but $L \ll N^2$. The result will include dependence in $N$ and on higher moments of the degree distribution.

\subsection{Cycles in two layers}

Now we outline the derivation for general cycles in a two layer multiplex. We proceed in the same way, but now there are three sets of nodes $\{ n^{(1)}_{q_1,q_2}  \}$ for the nodes connecting two edges in layer $1$, $\{ n^{(2)}_{q_1,q_2}  \}$ for two edges in layer $2$, and 
$\{ n^{(s)}_{q_1,q_2}  \}$ for the switches.
We sum over all possible such sets, and use three delta functions to count only those whose total number of nodes (of all degrees) match $s_{ij}$. 
The combinatorial factor for the number of ways to select the three sets is 
\begin{equation}
	B_{n_1,n_2,n_3}^{N} = \frac{N!}{n_1!n_2!n_3!(N-n_1-n_2-n_3)!}.
\end{equation}
There are $q_1(q_1-1)$ ways to select the two edges connecting each of the $n^{(1)}_{q_1,q_2}$ nodes, and similarly for $n^{(2)}_{q_1,q_2}$, while for the $n^{(s)}_{q_1,q_2}$ switches the factor is $q_1q_2$.
These considerations give us
\begin{multline}
	R(N,\l,P(\q)) =  \sum_{\{ n^{(1)}_{q_1,q_2}  \}} 
\delta\left( \sum_{q_1,q_2} n^{(1)}_{q_1,q_2} - s_{11} \right)  \\
	\sum_{\{ n^{(2)}_{q_1,q_2}  \}} 
	\delta\left( \sum_{q_1,q_2} n^{(2)}_{q_1,q_2} - s_{22} \right)  
	\sum_{\{ n^{(s)}_{q_1,q_2}  \}} 
	\delta\left( \sum_{q_1,q_2} n^{(s)}_{q_1,q_2} - s_{12} \right) \\
	 \prod_{q_1,q_2}\!\! B_{n^{(1)}_{q_1,q_2},n^{(2)}_{q_1,q_2},n^{(s)}_{q_1,q_2}}^{N_{q_1,q_2}} \!\!\!\! 
	\left[q_1(q_1-1)\right]^{n^{(1)}_{q_1,q_2}}\\
	 \times\left[q_2(q_2-1)\right]^{n^{(2)}_{q_1,q_2}} \left[q_1q_2\right]^{n^{(s)}_{q_1,q_2}}. \label{R_main}
\end{multline}
We can represent the delta function as an integral. Replacing the sums over sets $\{n^{(j)}_{q_1,q_2}\}$ with sums over $n^{(j)}_{q_1,q_2}$ after the product,
\begin{multline}
	R(N,\l,P(\q)) = \frac{N^{L}}{(2\pi)^3} \int \d x_1\d x_2\d x_3\\ 
	\exp\big\{ix_1(\sum_{q_1,q_2}\!n^{(1)}_{q_1,q_2}\! -\! s_{11})+ 
	ix_2(\sum_{q_1,q_2} \! n^{(2)}_{q_1,q_2} \!-\! s_{22})+\\
	ix_3(\sum_{q_1,q_2}\! n^{(2)}_{q_1,q_2} \!-\! s_{12})\big\}\times\\
	\prod_{q_1,q_2} 
	\sum_{n^{(1)}_{q_1,q_2} = 0}^{N_{q_1,q_2}}
	\!\!\!\sum_{n^{(2)}_{q_1,q_2} = 0}^{N_{q_1,q_2}\!-n^{(1)}_{q_1,q_2}}
	\sum_{n^{(s)}_{q_1,q_2} = 0}^{N_{q_1,q_2}\!-n^{(1)}_{q_1,q_2}\!-n^{(2)}_{q_1,q_2}}
	\!\!\!\!\!\!\!\!
	B_{n^{(1)}_{q_1,q_2},n^{(2)}_{q_1,q_2},n^{(s)}_{q_1,q_2}}^{N_{q_1,q_2}} \\
	\!\left[\frac{q_1(q_1-1)}{N}\right]^{n^{(1)}_{q_1,q_2}} 
	\!\left[\frac{q_2(q_2-1)}{N}\right]^{n^{(2)}_{q_1,q_2}} 
	\!\left[\frac{q_1q_2}{N}\right]^{n^{(s)}_{q_1,q_2}}.
\end{multline}
Each sum can be completed succesively using the binomial identity. Finally converting the product into a sum within an exponential, and writing it as an expectation value
we find
\begin{equation}
	R(N,\l,P(\q)) = \frac{N^{L}}{(2\pi)^3} \int
	\mathrm{d} x_1\,\mathrm{d} x_2\mathrm{d}\, x_3
	\exp \{ N\phi({\bf{x}}) \}. \label{R_full}
\end{equation}
where
\begin{multline}
\phi({\bf{x}})  = \frac{1}{N}\left[
\log\left(\frac{y_1}{N}\right)^{s_{11}}+\log\left(\frac{y_2}{N}\right)^{s_{22}}+\log\left(\frac{y_3}{N}\right)^{s_{12}}\right] +\\
	 \bigg\langle\log\left[ 1+\frac{b_1}{y_1} +\frac{b_2}{y_2} +\frac{b_3}{y_3} \right]\bigg\rangle
\end{multline}
and $y_j = Ne^{-ix_j}$ and $b_1 = q_1(q_1-1)$, $b_2 = q_2(q_2-1)$, and $b_3 = q_1q_2$.

As before, we expand $\phi({\bf{x}})$ about the local maximum $\bf{x}^*$ which is the 
simultaneous solution of
\begin{align}
\frac{s_{11}}{N} &= \left\langle \frac{b_1}{y_1^*[1+b_1/y_1^*+b_2/y_2^*+b_3/y_3^* ]} \right\rangle\label{star1}\\
\frac{s_{22}}{N} &= \left\langle \frac{b_2}{y_2^*[1+b_1/y_1^*+b_2/y_2^*+b_3/y_3^* ]} \right\rangle\label{star2}\\
\frac{s_{12}}{N} &= \left\langle \frac{b_3}{y_3^*[1+b_1/y_1^*+b_2/y_2^*+b_3/y_3^* ]} \right\rangle.\label{star3}
\end{align}
The Taylor expansion requires the second order derivatives:
\begin{align}
\phi_{jj} \equiv \frac{\partial^2 \phi}{\partial x_j^{2} }\Big|_{\bf{x}^*}&= - \frac{1}{y_j^*}
\left\langle  
\frac{b_j (1+b_k/y_k^* + b_l/y_l^*)}{(1+b_1/y_1^*+b_2/y_2^*+b_3/y_3^* )^2} 
\right\rangle \label{phi_jj}\\[5pt]
\phi_{jk} \equiv \frac{\partial^2 \phi}{\partial x_jx_k}\Big|_{\bf{x}^*} &= \frac{1}{y_j^*y_k^*}
\left\langle
\frac{b_jb_k}{(1+b_1/y_1^*+b_2/y_2^*+b_3/y_3^* )^2}
\right\rangle \label{phi_jk}
\end{align}
where the subscripts $j,k,l$ each take one of the values $1,2,3$.
Substituting the second order Taylor expansion of $\phi(\bf{x})$ around $\bf{x}^*$ into Eq. (\ref{R_full}) yields a more complex integral than in the single layer case, but its evaluation is still straightforward,
\begin{multline}
R(N,\l,P(\q)) =
\frac{N^{L} e^{N\phi({\bf{x}}^*)}}{(2\pi N)^{3/2} }
\left[
\phi_{11}\phi_{22}\phi_{33} \right.\\
\left. -\phi_{11}\phi_{23}^2-\phi_{22}\phi_{13}^2-\phi_{33}\phi_{12}^2
-2\phi_{12}\phi_{13}\phi_{23}
\right]^{-1/2}.\label{R_full_final}
\end{multline}
To evaluate $R(N,\l,P(\q))$ for a given distribution, one solves Eqs. (\ref{star1})-(\ref{star3}), evaluates Eqs. (\ref{phi_jj}) and (\ref{phi_jk}) and substitutes into Eq. (\ref{R_full_final})).

For $L\ll N$ we can neglect $\mathcal{O}(1/N)$, giving $y_1^* = N\langle b_1\rangle/s_{11}$ and similar expressions for $y_2^*$ and $y_3^*$. The cross derivatives $\phi_{12}$, $\phi_{13}$, and $\phi_{23}$ vanish, and we recover Eq. (\ref{R_short}). For larger $L$, dependence on $N$ will remain, and there will be dependence on higher moments of the degree distribution.

\bibliography{multiplex-cycles}

\begin{thebibliography}{25}%
\makeatletter
\providecommand \@ifxundefined [1]{%
 \@ifx{#1\undefined}
}%
\providecommand \@ifnum [1]{%
 \ifnum #1\expandafter \@firstoftwo
 \else \expandafter \@secondoftwo
 \fi
}%
\providecommand \@ifx [1]{%
 \ifx #1\expandafter \@firstoftwo
 \else \expandafter \@secondoftwo
 \fi
}%
\providecommand \natexlab [1]{#1}%
\providecommand \enquote  [1]{``#1''}%
\providecommand \bibnamefont  [1]{#1}%
\providecommand \bibfnamefont [1]{#1}%
\providecommand \citenamefont [1]{#1}%
\providecommand \href@noop [0]{\@secondoftwo}%
\providecommand \href [0]{\begingroup \@sanitize@url \@href}%
\providecommand \@href[1]{\@@startlink{#1}\@@href}%
\providecommand \@@href[1]{\endgroup#1\@@endlink}%
\providecommand \@sanitize@url [0]{\catcode `\\12\catcode `\$12\catcode
  `\&12\catcode `\#12\catcode `\^12\catcode `\_12\catcode `\%12\relax}%
\providecommand \@@startlink[1]{}%
\providecommand \@@endlink[0]{}%
\providecommand \url  [0]{\begingroup\@sanitize@url \@url }%
\providecommand \@url [1]{\endgroup\@href {#1}{\urlprefix }}%
\providecommand \urlprefix  [0]{URL }%
\providecommand \Eprint [0]{\href }%
\providecommand \doibase [0]{http://dx.doi.org/}%
\providecommand \selectlanguage [0]{\@gobble}%
\providecommand \bibinfo  [0]{\@secondoftwo}%
\providecommand \bibfield  [0]{\@secondoftwo}%
\providecommand \translation [1]{[#1]}%
\providecommand \BibitemOpen [0]{}%
\providecommand \bibitemStop [0]{}%
\providecommand \bibitemNoStop [0]{.\EOS\space}%
\providecommand \EOS [0]{\spacefactor3000\relax}%
\providecommand \BibitemShut  [1]{\csname bibitem#1\endcsname}%
\let\auto@bib@innerbib\@empty
\bibitem [{\citenamefont {Rinaldi}\ \emph {et~al.}(2001)\citenamefont
  {Rinaldi}, \citenamefont {Peerenboom},\ and\ \citenamefont
  {Kelly}}]{rinaldi2001identifying}%
  \BibitemOpen
  \bibfield  {author} {\bibinfo {author} {\bibfnamefont {S.~M.}\ \bibnamefont
  {Rinaldi}}, \bibinfo {author} {\bibfnamefont {J.~P.}\ \bibnamefont
  {Peerenboom}}, \ and\ \bibinfo {author} {\bibfnamefont {T.~K.}\ \bibnamefont
  {Kelly}},\ }\bibfield  {title} {\enquote {\bibinfo {title} {Identifying,
  understanding, and analyzing critical infrastructure interdependencies},}\
  }\href@noop {} {\bibfield  {journal} {\bibinfo  {journal} {IEEE Control Syst.
  Mag.}\ }\textbf {\bibinfo {volume} {21}},\ \bibinfo {pages} {11--25}
  (\bibinfo {year} {2001})}\BibitemShut {NoStop}%
\bibitem [{\citenamefont {Huang}\ \emph {et~al.}(2013)\citenamefont {Huang},
  \citenamefont {Vodenska}, \citenamefont {Havlin},\ and\ \citenamefont
  {Stanley}}]{huang2013cascading}%
  \BibitemOpen
  \bibfield  {author} {\bibinfo {author} {\bibfnamefont {X.}~\bibnamefont
  {Huang}}, \bibinfo {author} {\bibfnamefont {I.}~\bibnamefont {Vodenska}},
  \bibinfo {author} {\bibfnamefont {S.}~\bibnamefont {Havlin}}, \ and\ \bibinfo
  {author} {\bibfnamefont {H.~E.}\ \bibnamefont {Stanley}},\ }\href@noop {}
  {\bibfield  {journal} {\bibinfo  {journal} {Sci. Rep.}\ }\textbf {\bibinfo
  {volume} {3}},\ \bibinfo {pages} {1219} (\bibinfo {year} {2013})}\BibitemShut
  {NoStop}%
\bibitem [{\citenamefont {De~Domenico}\ \emph {et~al.}(2014)\citenamefont
  {De~Domenico}, \citenamefont {Sol\'{e}-Ribalta}, \citenamefont {G\'{o}mez},\
  and\ \citenamefont {Arenas}}]{dedomenico2014}%
  \BibitemOpen
  \bibfield  {author} {\bibinfo {author} {\bibfnamefont {M.}~\bibnamefont
  {De~Domenico}}, \bibinfo {author} {\bibfnamefont {A.}~\bibnamefont
  {Sol\'{e}-Ribalta}}, \bibinfo {author} {\bibfnamefont {S.}~\bibnamefont
  {G\'{o}mez}}, \ and\ \bibinfo {author} {\bibfnamefont {A.}~\bibnamefont
  {Arenas}},\ }\bibfield  {title} {\enquote {\bibinfo {title} {Navigability of
  interconnected networks under random failures},}\ }\href {\doibase
  10.1073/pnas.1318469111} {\bibfield  {journal} {\bibinfo  {journal} {Proc.
  Natl. Acad. Sci. U.S.A.}\ }\textbf {\bibinfo {volume} {111}},\ \bibinfo
  {pages} {8351--8356} (\bibinfo {year} {2014})},\ \Eprint
  {http://arxiv.org/abs/1306.0519} {arXiv:1306.0519} \BibitemShut {NoStop}%
\bibitem [{\citenamefont {Pocock}\ \emph {et~al.}(2012)\citenamefont {Pocock},
  \citenamefont {Evans},\ and\ \citenamefont {Memmott}}]{pocock2012robustness}%
  \BibitemOpen
  \bibfield  {author} {\bibinfo {author} {\bibfnamefont {M.~J.~O.}\
  \bibnamefont {Pocock}}, \bibinfo {author} {\bibfnamefont {D.~M.}\
  \bibnamefont {Evans}}, \ and\ \bibinfo {author} {\bibfnamefont
  {J.}~\bibnamefont {Memmott}},\ }\bibfield  {title} {\enquote {\bibinfo
  {title} {The robustness and restoration of a network of ecological
  networks},}\ }\href@noop {} {\bibfield  {journal} {\bibinfo  {journal}
  {Science}\ }\textbf {\bibinfo {volume} {335}},\ \bibinfo {pages} {973--976}
  (\bibinfo {year} {2012})}\BibitemShut {NoStop}%
\bibitem [{\citenamefont {Boccaletti}\ \emph {et~al.}(2014)\citenamefont
  {Boccaletti}, \citenamefont {Bianconi}, \citenamefont {Criado}, \citenamefont
  {Del~Genio}, \citenamefont {G{\'o}mez-Garde{\~n}es}, \citenamefont {Romance},
  \citenamefont {Sendi{\~n}a-Nadal}, \citenamefont {Wang},\ and\ \citenamefont
  {Zanin}}]{boccaletti2014structure}%
  \BibitemOpen
  \bibfield  {author} {\bibinfo {author} {\bibfnamefont {S.}~\bibnamefont
  {Boccaletti}}, \bibinfo {author} {\bibfnamefont {G.}~\bibnamefont
  {Bianconi}}, \bibinfo {author} {\bibfnamefont {R.}~\bibnamefont {Criado}},
  \bibinfo {author} {\bibfnamefont {C.~I.}\ \bibnamefont {Del~Genio}}, \bibinfo
  {author} {\bibfnamefont {J.}~\bibnamefont {G{\'o}mez-Garde{\~n}es}}, \bibinfo
  {author} {\bibfnamefont {M.}~\bibnamefont {Romance}}, \bibinfo {author}
  {\bibfnamefont {I.}~\bibnamefont {Sendi{\~n}a-Nadal}}, \bibinfo {author}
  {\bibfnamefont {Z.}~\bibnamefont {Wang}}, \ and\ \bibinfo {author}
  {\bibfnamefont {M.}~\bibnamefont {Zanin}},\ }\bibfield  {title} {\enquote
  {\bibinfo {title} {The structure and dynamics of multilayer networks},}\
  }\href@noop {} {\bibfield  {journal} {\bibinfo  {journal} {Phys. Rep.}\
  }\textbf {\bibinfo {volume} {544}},\ \bibinfo {pages} {1--122} (\bibinfo
  {year} {2014})}\BibitemShut {NoStop}%
\bibitem [{\citenamefont {Kivela}\ \emph {et~al.}(2014)\citenamefont {Kivela},
  \citenamefont {Arenas}, \citenamefont {Barthelemy}, \citenamefont {Gleeson},
  \citenamefont {Moreno},\ and\ \citenamefont {Porter}}]{kivela2014}%
  \BibitemOpen
  \bibfield  {author} {\bibinfo {author} {\bibfnamefont {M.}~\bibnamefont
  {Kivela}}, \bibinfo {author} {\bibfnamefont {A.}~\bibnamefont {Arenas}},
  \bibinfo {author} {\bibfnamefont {M.}~\bibnamefont {Barthelemy}}, \bibinfo
  {author} {\bibfnamefont {J.~P.}\ \bibnamefont {Gleeson}}, \bibinfo {author}
  {\bibfnamefont {Y.}~\bibnamefont {Moreno}}, \ and\ \bibinfo {author}
  {\bibfnamefont {M.~A.}\ \bibnamefont {Porter}},\ }\bibfield  {title}
  {\enquote {\bibinfo {title} {Multilayer networks},}\ }\href {\doibase
  10.1093/comnet/cnu016} {\bibfield  {journal} {\bibinfo  {journal} {Journal of
  Complex Networks}\ }\textbf {\bibinfo {volume} {2}},\ \bibinfo {pages}
  {203--271} (\bibinfo {year} {2014})}\BibitemShut {NoStop}%
\bibitem [{\citenamefont {De~Domenico}\ \emph {et~al.}(2016)\citenamefont
  {De~Domenico}, \citenamefont {Granell}, \citenamefont {Porter},\ and\
  \citenamefont {Arenas}}]{dedomenico2016physics}%
  \BibitemOpen
  \bibfield  {author} {\bibinfo {author} {\bibfnamefont {M.}~\bibnamefont
  {De~Domenico}}, \bibinfo {author} {\bibfnamefont {C.}~\bibnamefont
  {Granell}}, \bibinfo {author} {\bibfnamefont {M.~A.}\ \bibnamefont {Porter}},
  \ and\ \bibinfo {author} {\bibfnamefont {A.}~\bibnamefont {Arenas}},\
  }\bibfield  {title} {\enquote {\bibinfo {title} {The physics of multilayer
  networks},}\ }\href@noop {} {\bibfield  {journal} {\bibinfo  {journal} {arXiv
  preprint arXiv:1604.02021}\ } (\bibinfo {year} {2016})}\BibitemShut {NoStop}%
\bibitem [{\citenamefont {Baxter}\ \emph {et~al.}(2012)\citenamefont {Baxter},
  \citenamefont {Dorogovtsev}, \citenamefont {Goltsev},\ and\ \citenamefont
  {Mendes}}]{baxter2012avalanche}%
  \BibitemOpen
  \bibfield  {author} {\bibinfo {author} {\bibfnamefont {G.~J.}\ \bibnamefont
  {Baxter}}, \bibinfo {author} {\bibfnamefont {S.~N.}\ \bibnamefont
  {Dorogovtsev}}, \bibinfo {author} {\bibfnamefont {A.~V.}\ \bibnamefont
  {Goltsev}}, \ and\ \bibinfo {author} {\bibfnamefont {J.~F.F.}\ \bibnamefont
  {Mendes}},\ }\bibfield  {title} {\enquote {\bibinfo {title} {Avalanche
  collapse of interdependent networks},}\ }\href@noop {} {\bibfield  {journal}
  {\bibinfo  {journal} {Phys.Rev. Lett.}\ }\textbf {\bibinfo {volume} {109}},\
  \bibinfo {pages} {248701} (\bibinfo {year} {2012})}\BibitemShut {NoStop}%
\bibitem [{\citenamefont {Baxter}\ \emph {et~al.}(2016)\citenamefont {Baxter},
  \citenamefont {Cellai}, \citenamefont {Dorogovtsev}, \citenamefont
  {Goltsev},\ and\ \citenamefont {Mendes}}]{baxter2015unified}%
  \BibitemOpen
  \bibfield  {author} {\bibinfo {author} {\bibfnamefont {G.~J.}\ \bibnamefont
  {Baxter}}, \bibinfo {author} {\bibfnamefont {D.}~\bibnamefont {Cellai}},
  \bibinfo {author} {\bibfnamefont {S.~N.}\ \bibnamefont {Dorogovtsev}},
  \bibinfo {author} {\bibfnamefont {A.~V.}\ \bibnamefont {Goltsev}}, \ and\
  \bibinfo {author} {\bibfnamefont {J.~F.~F.}\ \bibnamefont {Mendes}},\
  }\bibfield  {title} {\enquote {\bibinfo {title} {A unified approach to
  percolation processes on multiplex networks},}\ }in\ \href@noop {} {\emph
  {\bibinfo {booktitle} {Interconnected Networks}}}\ (\bibinfo  {publisher}
  {Springer},\ \bibinfo {year} {2016})\ pp.\ \bibinfo {pages}
  {101--123}\BibitemShut {NoStop}%
\bibitem [{\citenamefont {Cozzo}\ \emph {et~al.}(2015)\citenamefont {Cozzo},
  \citenamefont {Kivel{\"a}}, \citenamefont {De~Domenico}, \citenamefont
  {Sol{\'e}-Ribalta}, \citenamefont {Arenas}, \citenamefont {G{\'o}mez},
  \citenamefont {Porter},\ and\ \citenamefont {Moreno}}]{cozzo2015structure}%
  \BibitemOpen
  \bibfield  {author} {\bibinfo {author} {\bibfnamefont {E.}~\bibnamefont
  {Cozzo}}, \bibinfo {author} {\bibfnamefont {M.}~\bibnamefont {Kivel{\"a}}},
  \bibinfo {author} {\bibfnamefont {M.}~\bibnamefont {De~Domenico}}, \bibinfo
  {author} {\bibfnamefont {A.}~\bibnamefont {Sol{\'e}-Ribalta}}, \bibinfo
  {author} {\bibfnamefont {A.}~\bibnamefont {Arenas}}, \bibinfo {author}
  {\bibfnamefont {S.}~\bibnamefont {G{\'o}mez}}, \bibinfo {author}
  {\bibfnamefont {M.~A.}\ \bibnamefont {Porter}}, \ and\ \bibinfo {author}
  {\bibfnamefont {Y.}~\bibnamefont {Moreno}},\ }\bibfield  {title} {\enquote
  {\bibinfo {title} {Structure of triadic relations in multiplex networks},}\
  }\href@noop {} {\bibfield  {journal} {\bibinfo  {journal} {New J. Phys.}\
  }\textbf {\bibinfo {volume} {17}},\ \bibinfo {pages} {073029} (\bibinfo
  {year} {2015})}\BibitemShut {NoStop}%
\bibitem [{\citenamefont {Newman}(2003)}]{newman2003}%
  \BibitemOpen
  \bibfield  {author} {\bibinfo {author} {\bibfnamefont {M.~E.~J.}\
  \bibnamefont {Newman}},\ }\bibfield  {title} {\enquote {\bibinfo {title} {The
  structure and function of complex networks},}\ }\href {\doibase
  10.1137/S003614450342480} {\bibfield  {journal} {\bibinfo  {journal} {SIAM
  Review}\ }\textbf {\bibinfo {volume} {45}},\ \bibinfo {pages} {167--256}
  (\bibinfo {year} {2003})},\ \Eprint {http://arxiv.org/abs/cond-mat/0303516}
  {arXiv:cond-mat/0303516} \BibitemShut {NoStop}%
\bibitem [{\citenamefont {Bianconi}\ and\ \citenamefont
  {Capocci}(2003)}]{bianconi2003number}%
  \BibitemOpen
  \bibfield  {author} {\bibinfo {author} {\bibfnamefont {G.}~\bibnamefont
  {Bianconi}}\ and\ \bibinfo {author} {\bibfnamefont {A.}~\bibnamefont
  {Capocci}},\ }\bibfield  {title} {\enquote {\bibinfo {title} {Number of loops
  of size $h$ in growing scale-free networks},}\ }\href@noop {} {\bibfield
  {journal} {\bibinfo  {journal} {Phys. Rev. Lett.}\ }\textbf {\bibinfo
  {volume} {90}},\ \bibinfo {pages} {078701} (\bibinfo {year}
  {2003})}\BibitemShut {NoStop}%
\bibitem [{\citenamefont {Bollob{\'a}s}\ and\ \citenamefont
  {Riordan}(2003)}]{bollobas2003mathematical}%
  \BibitemOpen
  \bibfield  {author} {\bibinfo {author} {\bibfnamefont {B.}~\bibnamefont
  {Bollob{\'a}s}}\ and\ \bibinfo {author} {\bibfnamefont {O.~M.}\ \bibnamefont
  {Riordan}},\ }\bibfield  {title} {\enquote {\bibinfo {title} {Mathematical
  results on scale-free random graphs},}\ }in\ \href@noop {} {\emph {\bibinfo
  {booktitle} {Handbook of Graphs and Networks: From the Genome to the
  Internet}}},\ \bibinfo {editor} {edited by\ \bibinfo {editor} {\bibfnamefont
  {S.}~\bibnamefont {Bornholdt}}\ and\ \bibinfo {editor} {\bibfnamefont
  {H.~G.}\ \bibnamefont {Schuster}}}\ (\bibinfo  {publisher} {Wiley-VCH},\
  \bibinfo {address} {Weinheim},\ \bibinfo {year} {2003})\ pp.\ \bibinfo
  {pages} {1--34}\BibitemShut {NoStop}%
\bibitem [{\citenamefont {Melnik}\ \emph {et~al.}(2011)\citenamefont {Melnik},
  \citenamefont {Hackett}, \citenamefont {Porter}, \citenamefont {Mucha},\ and\
  \citenamefont {Gleeson}}]{melnik2011}%
  \BibitemOpen
  \bibfield  {author} {\bibinfo {author} {\bibfnamefont {S.}~\bibnamefont
  {Melnik}}, \bibinfo {author} {\bibfnamefont {A.}~\bibnamefont {Hackett}},
  \bibinfo {author} {\bibfnamefont {M.~A.}\ \bibnamefont {Porter}}, \bibinfo
  {author} {\bibfnamefont {P.}~\bibnamefont {Mucha}}, \ and\ \bibinfo {author}
  {\bibfnamefont {J.}~\bibnamefont {Gleeson}},\ }\bibfield  {title} {\enquote
  {\bibinfo {title} {The unreasonable effectiveness of tree-based theory for
  networks with clustering},}\ }\href {\doibase 10.1103/PhysRevE.83.036112}
  {\bibfield  {journal} {\bibinfo  {journal} {Phys. Rev. E}\ }\textbf {\bibinfo
  {volume} {83}},\ \bibinfo {pages} {036112} (\bibinfo {year}
  {2011})}\BibitemShut {NoStop}%
\bibitem [{\citenamefont {Kim}\ and\ \citenamefont {Goh}(2013)}]{kim2013}%
  \BibitemOpen
  \bibfield  {author} {\bibinfo {author} {\bibfnamefont {J.~Y.}\ \bibnamefont
  {Kim}}\ and\ \bibinfo {author} {\bibfnamefont {K.}~\bibnamefont {Goh}},\
  }\bibfield  {title} {\enquote {\bibinfo {title} {Coevolution and correlated
  multiplexity in multiplex networks},}\ }\href {\doibase
  10.1103/PhysRevLett.111.058702} {\bibfield  {journal} {\bibinfo  {journal}
  {Phys. Rev. Lett.}\ }\textbf {\bibinfo {volume} {111}},\ \bibinfo {pages}
  {058702} (\bibinfo {year} {2013})}\BibitemShut {NoStop}%
\bibitem [{\citenamefont {Hackett}\ \emph {et~al.}(2016)\citenamefont
  {Hackett}, \citenamefont {Cellai}, \citenamefont {G{\'o}mez}, \citenamefont
  {Arenas},\ and\ \citenamefont {Gleeson}}]{hackett2016}%
  \BibitemOpen
  \bibfield  {author} {\bibinfo {author} {\bibfnamefont {A.}~\bibnamefont
  {Hackett}}, \bibinfo {author} {\bibfnamefont {D.}~\bibnamefont {Cellai}},
  \bibinfo {author} {\bibfnamefont {S.}~\bibnamefont {G{\'o}mez}}, \bibinfo
  {author} {\bibfnamefont {A.}~\bibnamefont {Arenas}}, \ and\ \bibinfo {author}
  {\bibfnamefont {J.~P.}\ \bibnamefont {Gleeson}},\ }\bibfield  {title}
  {\enquote {\bibinfo {title} {Bond percolation on multiplex networks},}\
  }\href@noop {} {\bibfield  {journal} {\bibinfo  {journal} {Physical Review
  X}\ }\textbf {\bibinfo {volume} {6}},\ \bibinfo {pages} {021002} (\bibinfo
  {year} {2016})}\BibitemShut {NoStop}%
\bibitem [{\citenamefont {Bollob\'{a}s}(2001)}]{bollobas2001book}%
  \BibitemOpen
  \bibfield  {author} {\bibinfo {author} {\bibfnamefont {B\'{e}la}\
  \bibnamefont {Bollob\'{a}s}},\ }\href {\doibase 10.1017/CBO9780511814068}
  {\emph {\bibinfo {title} {Random Graphs}}},\ \bibinfo {edition} {2nd}\ ed.\
  (\bibinfo  {publisher} {Cambridge University Press},\ \bibinfo {address}
  {Cambridge},\ \bibinfo {year} {2001})\BibitemShut {NoStop}%
\bibitem [{\citenamefont {Harary}\ and\ \citenamefont
  {Manvel}(1971)}]{harary1971}%
  \BibitemOpen
  \bibfield  {author} {\bibinfo {author} {\bibfnamefont {F.}~\bibnamefont
  {Harary}}\ and\ \bibinfo {author} {\bibfnamefont {B.}~\bibnamefont
  {Manvel}},\ }\bibfield  {title} {\enquote {\bibinfo {title} {On the number of
  cycles in a graph},}\ }\href@noop {} {\bibfield  {journal} {\bibinfo
  {journal} {Math. Slovaca}\ }\textbf {\bibinfo {volume} {21}},\ \bibinfo
  {pages} {55--63} (\bibinfo {year} {1971})}\BibitemShut {NoStop}%
\bibitem [{\citenamefont {Abouelaoualim}\ \emph {et~al.}(2010)\citenamefont
  {Abouelaoualim}, \citenamefont {Das}, \citenamefont {de~la Vega},
  \citenamefont {Karpinski}, \citenamefont {Manoussakis}, \citenamefont
  {Martinhon},\ and\ \citenamefont {Saad}}]{abouelaoualim2010}%
  \BibitemOpen
  \bibfield  {author} {\bibinfo {author} {\bibfnamefont {A.}~\bibnamefont
  {Abouelaoualim}}, \bibinfo {author} {\bibnamefont {Das}}, \bibinfo {author}
  {\bibfnamefont {W.~Fernandez}\ \bibnamefont {de~la Vega}}, \bibinfo {author}
  {\bibfnamefont {M.}~\bibnamefont {Karpinski}}, \bibinfo {author}
  {\bibfnamefont {Y.}~\bibnamefont {Manoussakis}}, \bibinfo {author}
  {\bibfnamefont {C.~A.}\ \bibnamefont {Martinhon}}, \ and\ \bibinfo {author}
  {\bibfnamefont {R.}~\bibnamefont {Saad}},\ }\bibfield  {title} {\enquote
  {\bibinfo {title} {Cycles and paths in edge-colored graphs with given
  degrees},}\ }\href {\doibase 10.1002/jgt.20440} {\bibfield  {journal}
  {\bibinfo  {journal} {J. Graph Theory}\ }\textbf {\bibinfo {volume} {64}},\
  \bibinfo {pages} {63--86} (\bibinfo {year} {2010})}\BibitemShut {NoStop}%
\bibitem [{\citenamefont {Wang}\ and\ \citenamefont
  {Desmedt}(2011)}]{wang2011}%
  \BibitemOpen
  \bibfield  {author} {\bibinfo {author} {\bibfnamefont {Y.}~\bibnamefont
  {Wang}}\ and\ \bibinfo {author} {\bibfnamefont {Y.}~\bibnamefont {Desmedt}},\
  }\bibfield  {title} {\enquote {\bibinfo {title} {Edge-colored graphs with
  applications to homogeneous faults},}\ }\href {\doibase
  10.1016/j.ipl.2011.03.017} {\bibfield  {journal} {\bibinfo  {journal}
  {Inform. Process. Lett.}\ }\textbf {\bibinfo {volume} {111}},\ \bibinfo
  {pages} {634--641} (\bibinfo {year} {2011})}\BibitemShut {NoStop}%
\bibitem [{\citenamefont {Jain}(2004)}]{jain2004}%
  \BibitemOpen
  \bibfield  {author} {\bibinfo {author} {\bibfnamefont {K.}~\bibnamefont
  {Jain}},\ }\bibfield  {title} {\enquote {\bibinfo {title} {Security based on
  network topology against the wiretapping attack},}\ }\href {\doibase
  10.1109/mwc.2004.1269720} {\bibfield  {journal} {\bibinfo  {journal} {IEEE
  wirel. commun.}\ }\textbf {\bibinfo {volume} {11}},\ \bibinfo {pages}
  {68--71} (\bibinfo {year} {2004})}\BibitemShut {NoStop}%
\bibitem [{\citenamefont {Bender}\ and\ \citenamefont
  {Canfield}(1978)}]{bender1978asymptotic}%
  \BibitemOpen
  \bibfield  {author} {\bibinfo {author} {\bibfnamefont {E.~A.}\ \bibnamefont
  {Bender}}\ and\ \bibinfo {author} {\bibfnamefont {E.~R.}\ \bibnamefont
  {Canfield}},\ }\bibfield  {title} {\enquote {\bibinfo {title} {The asymptotic
  number of labeled graphs with given degree sequences},}\ }\href@noop {}
  {\bibfield  {journal} {\bibinfo  {journal} {J. Combinator. Theor. Series A}\
  }\textbf {\bibinfo {volume} {24}},\ \bibinfo {pages} {296} (\bibinfo {year}
  {1978})}\BibitemShut {NoStop}%
\bibitem [{\citenamefont {Bollob{\'a}s}(1980)}]{bollobas1980probabilistic}%
  \BibitemOpen
  \bibfield  {author} {\bibinfo {author} {\bibfnamefont {B.}~\bibnamefont
  {Bollob{\'a}s}},\ }\bibfield  {title} {\enquote {\bibinfo {title} {A
  probabilistic proof of an asymptotic formula for the number of labelled
  regular graphs},}\ }\href@noop {} {\bibfield  {journal} {\bibinfo  {journal}
  {Eur. J. Combinator.}\ }\textbf {\bibinfo {volume} {1}},\ \bibinfo {pages}
  {311} (\bibinfo {year} {1980})}\BibitemShut {NoStop}%
\bibitem [{\citenamefont {Bianconi}\ and\ \citenamefont
  {Marsili}(2005)}]{bianconi2005loops}%
  \BibitemOpen
  \bibfield  {author} {\bibinfo {author} {\bibfnamefont {G.}~\bibnamefont
  {Bianconi}}\ and\ \bibinfo {author} {\bibfnamefont {M.}~\bibnamefont
  {Marsili}},\ }\bibfield  {title} {\enquote {\bibinfo {title} {Loops of any
  size and hamilton cycles in random scale-free networks},}\ }\href@noop {}
  {\bibfield  {journal} {\bibinfo  {journal} {J. Stat. Mech. Theor. Exp.}\
  }\textbf {\bibinfo {volume} {2005}},\ \bibinfo {pages} {P06005} (\bibinfo
  {year} {2005})}\BibitemShut {NoStop}%
\bibitem [{\citenamefont {Bianconi}\ and\ \citenamefont
  {Marsili}(2006)}]{bianconi2006}%
  \BibitemOpen
  \bibfield  {author} {\bibinfo {author} {\bibfnamefont {G.}~\bibnamefont
  {Bianconi}}\ and\ \bibinfo {author} {\bibfnamefont {M.}~\bibnamefont
  {Marsili}},\ }\bibfield  {title} {\enquote {\bibinfo {title} {Effect of
  degree correlations on the loop structure of scale-free networks},}\ }\href
  {\doibase 10.1103/physreve.73.066127} {\bibfield  {journal} {\bibinfo
  {journal} {Phys. Rev. E}\ }\textbf {\bibinfo {volume} {73}},\ \bibinfo
  {pages} {066127+} (\bibinfo {year} {2006})}\BibitemShut {NoStop}%
\end{thebibliography}%

\end{document}